\documentclass[12pt,a4paper,onecolumn]{article}

\usepackage{graphicx}
\usepackage{wrapfig}

\newcommand{\gvec}[1]{\hbox{\boldmath$#1$\unboldmath}}

\begin{document}

\title{Effects of the nuclear correlations on the neutrino-oxygen interactions}
\author{J. Marteau} 
\date{\textit{Institut de Physique nucl\'eaire de Lyon,\\ 43 bld du 11 novembre 1918, 69622 Villeurbanne} }

\maketitle

\begin{abstract}
We perform a calculation of the absolute charged current neutrino-oxygen events 
rates relevant in the atmospheric neutrino experiments. The inclusive reaction 
cross-section is split into exclusive channels, which are classified according 
to the number of \v{C}erenkov rings they produce. The model includes the 
effects of residual interaction in a RPA scheme with both nucleon-hole and 
Delta-hole excited states and the effects of (\textit{np-nh}) excitations 
(\textit{n=2,3}). Our result is that although the flavor ratio 
$\mu/e$ remains almost unaffected by the nuclear effects considered here and 
often neglected in the Monte-Carlo simulations, the absolute events rates are 
subject to important modifications.
\end{abstract} 


\section{Introduction}     
     
Neutrino physics is among the hottest topics of particle physics with the recent
indications in favor of neutrino oscillations. After the solar neutrino 
deficit, the apparent anomaly in the ratio of muon to electron atmospheric 
neutrinos $ R_{\mu/e} = ( N_{\nu_\mu} + N_{\bar{\nu}_\mu} ) / ( N_{\nu_e} + 
N_{\bar{\nu}_e} ) $ observed by (Super-)Kamiokande \cite{kamiokande,superkamiokande}, IMB \cite{imb}, Soudan-2 \cite{soudan} and the asymmetry in the zenithal distributions of the $ \mu-\mathrm{type} $ events in Super-Kamiokande \cite{superkamiokande} have given a strong support to the oscillation hypothesis: $ \nu_\mu \longrightarrow \nu_x $ where 
$ \nu_x = \nu_\tau $ (\textit{i.e. active-active} transition) or $ \nu_x = 
\nu_s $ (\textit{i.e. active-sterile} transition). The solution of the atmospheric neutrinos anomaly in terms of $ \nu_\mu \longrightarrow \nu_e $ oscillations has been excluded by the Chooz collaboration \cite{chooz}.

A number of atmospheric neutrinos experiments use large underground water 
\v{C}erenkov detectors. In these experiments only "one \v{C}erenkov ring" (1 
\v{C}.R.) events are retained for the analysis. These events are usually assumed
to be produced by quasi-elastic charged current interactions in which a charged 
lepton is emitted above \v{C}erenkov threshold and leads to one \v{C}erenkov 
ring. The nucleon which is ejected from the nucleus is in general below 
threshold and therefore does not produce another ring.
The region of energy transfer in processes involving atmospheric neutrinos of 
$ \sim $ 1 GeV extends from the quasi-elastic peak to the Delta resonance 
region. The evaluation of the nuclear responses in the latter region usually 
relies on the assumption that the Delta decays into a pion and a nucleon (this 
is the case for example in ref. \cite{kim/schramm/horowitz} where the authors use a relativistic model \textit{\`a la} Walecka to compute the nuclear response functions). The pion 
leading to an additional \v{C}erenkov ring, this charged current event belongs 
to the two \v{C}erenkov rings (2 \v{C}.R.) class and is rejected by the 
experimental cuts. Thus theoretical calculations are often limited to the 
quasi-elastic peak which is treated in Fermi gas models or with more elaborate 
treatments taking into account the shell structure of the oxygen nucleus and 
RPA type correlations \cite{engel/vogel}. 

However the nuclear dynamics is far more complex than this simple picture. 
Indeed the pion in the nucleus is a quasi-particle with a broad width and can 
decay for instance into a \textit{particle-hole} excitation. Therefore the decay of a 
Delta in the nuclear medium can lead to a nucleon and a \textit{particle-hole} 
state. In such a process, two nucleons are ejected from the nucleus, none of 
them producing a \v{C}erenkov ring, and the event belongs to the 1 \v{C}.R. 
class. Furthermore (\textit{2p-2h}) states may also be directly excited in the 
nucleus without excitation of the Delta resonance. This process also results 
in the emission of two nucleons and the event belongs to the 1 \v{C}.R. class. 
Following these arguments, we perform a full calculation of the neutrino-oxygen 
cross sections beyond the quasi-elastic assumption, with the identification of 
the possible final states. This procedure leads to a complete evaluation of 
the 1 \v{C}.R. events yields in the atmospheric neutrinos experiments and its 
impact in the description of the retained neutrino events in the detectors has 
to be investigated.
 
The starting point of this calculation is the inclusive charged current cross 
section for the reaction \mbox{$ \nu_l \, (\bar{\nu}_l) + ^{16}\hbox{O} 
\longrightarrow l^- \, (l^+) + X $},
\begin{eqnarray} \label{eq:1}
\frac{\partial^2\sigma}{\partial\Omega \partial k^\prime} & = & \frac{G_F^2 \, 
\cos^2\theta_c \, (\gvec{k}^\prime)^2}{2 \, \pi^2} \, \cos^2\frac{\theta}{2} \, 
\left[ G_E^2 \, (\frac{q_\mu^2}{\gvec{q}^2})^2 \, R_\tau^{NN} \right. \nonumber 
\\ & + & G_A^2 \, \frac{( M_\Delta - M )^2}{2 \, \gvec{q}^2} \, R_{\sigma\tau (L)}^{N\Delta} +  G_A^2 \, \frac{( M_\Delta - M )^2}{\gvec{q}^2} R_{\sigma\tau (L)}^{\Delta\Delta} \nonumber \\ & + & 
( G_M^2 \, \frac{\omega^2}{\gvec{q}^2} + G_A^2 ) \, 
( - \frac{q_\mu^2}{\gvec{q}^2} + 2 \tan^2\frac{\theta}{2} ) \,
( R_{\sigma\tau (T)}^{NN} + 2 R_{\sigma\tau (T)}^{N\Delta} 
+ R_{\sigma\tau (T)}^{\Delta\Delta} ) \nonumber \\ 
& \pm & \left. 2 \, G_A \, G_M \, \frac{k + k^\prime}{M} \, 
\tan^2\frac{\theta}{2} \, 
( R_{\sigma\tau (T)}^{NN} + 2 R_{\sigma\tau (T)}^{N\Delta} 
+ R_{\sigma\tau (T)}^{\Delta\Delta} ) \right] 
\end{eqnarray}
where $ G_F $ is the weak coupling constant, $ \theta_c $ the Cabbibo angle, 
$ k $ and $ k^\prime $ the initial and final lepton momenta, $ q_\mu = k_\mu - 
k_\mu^\prime = ( \omega,\gvec{q} ) $ the four momentum transferred to the 
nucleus, $ \theta $ the scattering angle, $ M_\Delta $ ($ M $) the Delta 
(nucleon) mass. The plus (minus) sign in eq. (\ref{eq:1}) stands for the 
neutrino (antineutrino) case. In a provisional approximation, to be lifted after, we have 
neglected in eq. (\ref{eq:1}) the lepton masses and we have kept the leading 
terms in the development of the hadronic current in $ p/M $, where $ p $ 
denotes the initial nucleon momentum. The electric, magnetic and axial form 
factors are taken in the standard dipole parameterization with the following 
normalizations: $ G_E(0) = 1.0 $, $ G_M(0) = 4.71 $ and $ G_A(0) = 1.25 $. We 
have introduced the inclusive \textit{isospin} ($ R_\tau $), 
\textit{spin-isospin longitudinal} ($ R_{\sigma\tau (L)} $) and 
\textit{spin-isospin transverse} ($ R_{\sigma\tau (T)} $) nuclear responses 
functions (the longitudinal and transverse character of these last two 
responses refers to the direction of the spin operator with respect to the 
direction of the transferred momentum):
\begin{equation} \label{eq:2}
R_\alpha^{PP^\prime} = \sum_n \, 
\langle n | \sum_{j=1}^A \, O_\alpha^P(j) \, 
e^{ i \, \gvec{q}.\gvec{x}_j } | 0 \rangle 
\langle n | \sum_{k=1}^A \, O_\alpha^{P^\prime}(k) \, 
e^{ i \, \gvec{q}.\gvec{x}_k } | 0 \rangle^* \, \delta (\omega - E_n + E_0 )  
\end{equation}
where the operators have the following forms:
$$ 
O_\alpha^N(j) = \tau_j^\pm, \,\,\, ( \gvec{\sigma}_j . \widehat{q} ) \, 
\tau_j^\pm, \,\,\, (( \gvec{\sigma}_j \times \widehat{q} ) \times \widehat{q} ) 
\, \tau_j^\pm, 
$$
for $ \alpha = \tau $, $ \sigma\tau (L) $, $ \sigma\tau (T) $, and 
$$
O_\alpha^\Delta(j) = ( \gvec{S}_j . \widehat{q} ) \, T_j^\pm, \,\,\, (( 
\gvec{S}_j \times \widehat{q} ) \times \widehat{q} ) \, T_j^\pm, 
$$
for $ \alpha = \sigma\tau (L) $, $ \sigma\tau (T) $. 
In the above expressions, the superscript $ P $ ($ P = N $ or $ \Delta $) 
denotes the type of the \textit{Particle-hole} excitations 
(\textit{Nucleon-hole} or \textit{Delta-hole}) induced by the operator $ 
O_\alpha^P $. The operators $ S $ and $ T $ are the usual 1/2 to 3/2 
transition operators in the spin and isospin space (for instance see \cite{ericson/weise}). 
In this work we neglect 
the small quadrupole transition connecting the nucleon to the Delta through 
the pure isospin operator, therefore the isospin response just involves 
nucleon-hole excitations. Note that we have assumed the existence of a scaling 
law between the nucleon and Delta magnetic and axial form factors \cite{chew/low}: 
$$
 G_M^* / G_M = G_A^* / G_A = f^* / f, 
$$ where $ f^* $ ($ f $) is the $ \pi \, 
N \, \Delta $ 
($ \pi \, N\, N $) coupling constant. For a matter of convenience, we have 
incorporated the scaling factor $ f^* / f  = 2.2 $ into the responses. 

\section{Formalism}     
     
The evaluation of the nuclear responses is performed within the model 
developed by Delorme and Guichon for the interpretation of the 
($^3\hbox{He},t $) charge exchange experiments \cite{delorme/guichon,delorme/guichon/pl}. In this model the 
polarization propagators 
$ \Pi^0 (\omega,\gvec{q},\gvec{q}^\prime) $ without nuclear correlations are 
evaluated in a semi-classical approximation to properly take into account the 
finite size effects. This implies the use of a local Fermi momentum $ k_F(r) $ 
which is calculated by the means of an experimental nuclear density: 
$ k_F(r) = ( 3/2 \, \pi^2 \, \rho(r) )^{1/3} $. Note that this procedure 
differs a little from the pure semi-classical one \footnote{For a pure quantum approach in the low energy part of the nuclear response, applied in the context of terrestrial neutrinos experiments, see ref. \cite{giai,kolbe/langanke}} but it has been found to give better 
results for the $\pi-\mathrm{nucleus}$ reactions. The "bare" polarization propagators $ \Pi^0 $ (in the following 
"bare" will mean that the nuclear correlations are switched off) are then used 
as an input to exactly solve the RPA equations in the ring approximation, as 
we will develop in the following. In ref. \cite{delorme/guichon,delorme/guichon/pl} the authors gave satisfactory 
fits to the set of the experimental data. This model was also confronted to 
the pion-nuclei experimental results \cite{laktineh} and the agreement obtained for the 
total and elastic cross sections was remarkably good. 

\begin{figure}[ht]
\begin{center}
\includegraphics[width=12cm,height=7cm]{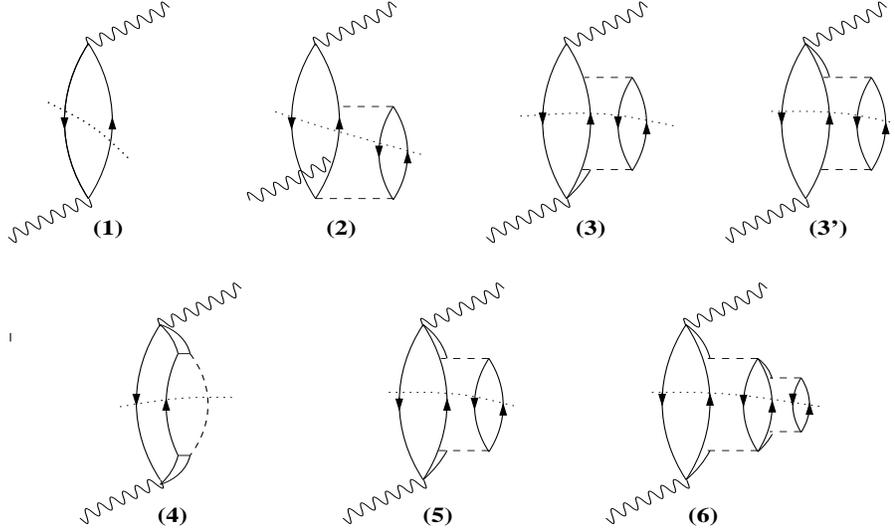}
\caption{\label{fig:1} \textit{Feynman graphs of the partial polarization 
propagators: $NN$ quasi-elastic (1), $NN$ (2p-2h) (2), $N\Delta$ (2p-2h) (3), 
$\Delta N$ (2p-2h) (3'), $\Delta\Delta$ ($\pi N$) (4), $\Delta\Delta$ (2p-2h) 
(5), $\Delta\Delta$ (3p-3h) (6). The conventions for the various lines drawings are given in the text.}}
\end{center} 
\end{figure}

As mentioned above, the first step of the calculation is the evaluation of the 
bare polarization propagators. A crucial ingredient of the model is the Delta 
resonance width modified by the nuclear effects. We adopt the parameterization 
of ref. \cite{oset/salcedo} where the Delta width is split into the contributions of different 
decay channels: the "quasi-elastic" channel, 
$ \Delta \longrightarrow \pi \, N $, modified by the Pauli blocking of the 
nucleon and the distortion of the pion, the two-body (\textit{2p-2h}) and 
three-body (\textit{3p-3h}) absorption channels. This parameterization leads to 
a good description of pion-nuclear reactions. At resonance we find a Delta 
width around 130 MeV, a value rather close to the free case. This value 
reflects the importance of the two- and three-body absorption channels which 
are large enough to counteract the effect of the Pauli blocking and lead to 
this overall enhancement of the Delta width. Note furthermore that at 
resonance the "quasi-elastic" channel modified by the medium effects, is 
almost equal to the free "quasi-elastic" one. 
The model of Delorme and Guichon also accounts for the (\textit{2p-2h}) 
excitations which are not reducible to a modified Delta width. The 
evaluation of such processes is performed by extrapolating the calculations of 
two-body pion absorption at threshold given in ref. \cite{shimizu/faessler}. We have limited 
ourselves to the imaginary part of these two-body polarization propagators, 
the comparison with experimental data such as pion-nucleus scattering or 
($ e,e^\prime $) scattering giving satisfactory results to that order of 
approximation.   
By construction, the bare polarization propagator 
$ \Pi^0(\omega,\gvec{q},\gvec{q}^\prime) $ is the sum of the following partial 
components:
\begin{enumerate}       
\item $ NN $ quasi-elastic (the standard Lindhard function), 
\item $ NN $ (\textit{2p-2h}), 
\item $ N\Delta $ and $ 3^\prime. $ $ \Delta N $ (\textit{2p-2h}), 
\item $ \Delta\Delta $ ($ \pi \, N $),
\item $ \Delta\Delta $ (\textit{2p-2h}),
\item $ \Delta\Delta $ (\textit{3p-3h}),
\end{enumerate}
where the notation $ N $ ($ \Delta $) stands for Nucleon-hole (Delta-hole) 
states as previously. The Feynman graphs corresponding to this partial 
polarization propagators are displayed on fig. (\ref{fig:1}) with the 
following conventions: the wiggled lines represent the external probe, the 
full lines correspond to the propagation of a nucleon (or a hole), the double 
lines to the propagation of a Delta, the dashed lines to an effective 
interaction between nucleons and/or Deltas. Finally the dotted lines indicate which 
intermediate state has to be placed on-shell to obtain the desired partial 
nuclear response. Note that in the case of (\textit{np-nh}) polarization 
propagators the number of graphs is large and we just give one example in the 
figure.

The bare responses are then given by the standard relations:
\begin{equation} \label{eq:3}
R^0(k)(\omega,q) = -\frac{1}{\pi} \, \mathrm{Im}(\Pi^0(k)(\omega,q,q) ),  
\end{equation}
with  the obvious sum rule: 
\begin{equation}
R^0(\omega,q) = -\frac{1}{\pi} \, \mathrm{Im}( \Pi^0(\omega,q,q) ) = \sum_{k=1}^{n_k} R^0(k)(\omega,q),
\end{equation}
where $ n_k $ denotes the number of partial reaction channels ( $ n_k = 7 $ in our model).

\begin{figure}[ht]
\begin{center}
\includegraphics[width=12cm,height=7cm]{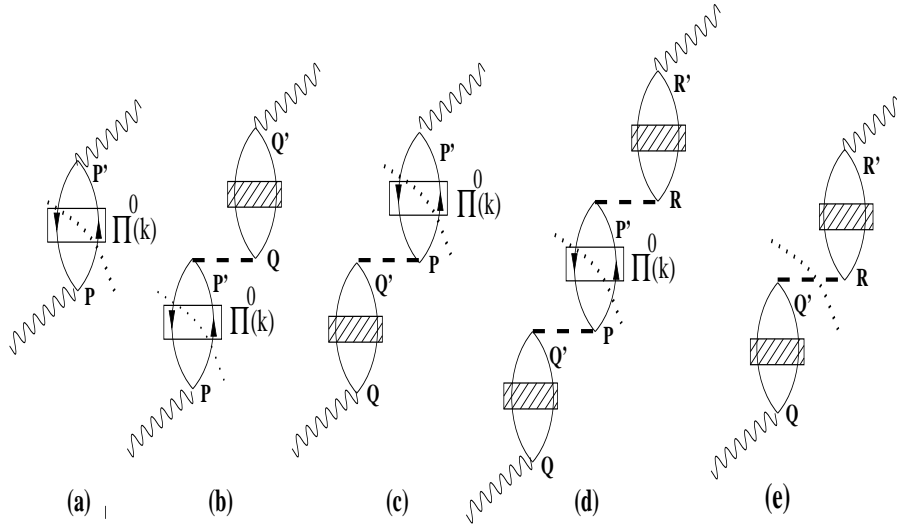}
\caption{\label{fig:2} \textit{Graphic representation of the partial RPA 
response functions. (a)-(d): incoherent partial response functions. (e): 
coherent partial response function. The hatched rings correspond to the 
inclusive RPA polarization propagator, the free rings to the partial bare 
polarization propagators, the dashed line to the effective interaction and the 
dotted line indicates which intermediate state is placed on-shell.}} 
\end{center}
\end{figure}

Following the method detailed in ref. \cite{delorme/guichon,delorme/guichon/pl} we include the effects of nuclear 
correlations by exactly solving the RPA equations in the ring approximation. 
For instance the inclusive RPA polarization propagators 
$ \Pi(\omega,\gvec{q},\gvec{q}^\prime) $ are solution of the generic equation:
\begin{equation} \label{eq:4}
\Pi = \Pi^0 + \Pi^0 \, V \Pi
\end{equation}
where $ V $ denotes the effective interaction between \textit{particle-hole} 
excitations and $ \Pi^0(\omega,\gvec{q},\gvec{q^\prime}) $ the inclusive bare polarization 
propagator calculated previously and used here as an input. In the spin-isospin channel the RPA equations couple the $ L,T $ and the $ N, \Delta $ components of the polarization propagators. For the effective interaction relevant in the isospin and spin-isospin channels, we 
use the standard $ \pi \, + \, \rho \, + \delta-\mathrm{function} $ 
parameterization:
\begin{eqnarray} \label{eq:5}
V_{NN} & = & ( f^\prime \, + \, V_\pi \, + \, V_\rho \, + \, V_{g^\prime} )  \,\, 
\gvec{\tau}_1 . \gvec{\tau}_2 \nonumber \\
V_{N \Delta} & = & ( V_\pi \, + \, V_\rho \, + \, V_{g^\prime} )  \,\, 
\gvec{\tau}_1 . \gvec{T}^\dagger_2 \nonumber \\  
V_{\Delta N} & = & ( V_\pi \, + \, V_\rho \, + \, V_{g^\prime} )  \,\, 
\gvec{T}_1 . \gvec{\tau}_2 \nonumber \\  
V_{\Delta\Delta} & = & ( V_\pi \, + \, V_\rho \, + \, V_{g^\prime} )  \,\, 
\gvec{T}_1 . \gvec{T}^\dagger_2  
\end{eqnarray}
where in the $ NN $ case, for example (the $ N \Delta $, $ \Delta N $ and $ \Delta\Delta $ cases are obtained with the appropriate replacements $ \sigma \longrightarrow S $):
\begin{eqnarray}
V_\pi & = & F_\pi^2 \,\, ( \frac{\gvec{q}^2}{\omega^2 - \gvec{q}^2 - m_\pi^2} ) \,\, \gvec{\sigma}_1 . \widehat{q} \,\, \gvec{\sigma}_2 . \widehat{q} \nonumber \\
V_\rho & = & F_\rho^2 \,\, ( \frac{\gvec{q}^2}{\omega^2 - \gvec{q}^2 - m_\rho^2} ) \,\, \gvec{\sigma}_1 \times \widehat{q} \,\, \gvec{\sigma}_2 \times \widehat{q} \nonumber \\
V_{g^\prime} & = & F_\pi^2 \,\,\, g^\prime \,\,\, \gvec{\sigma}_1 . \gvec{\sigma}_2 
\end{eqnarray}
In the preceding equations, $ F_\pi(q) $ and $ F_\rho(q) $ are the standard 
pion-nucleon and rho-nucleon form factors. The values we adopt for the relevant parameters can be found in ref. \cite{laktineh}. In particular we take the "common" Landau-Migdal parameters: $$ f^\prime = 0.6, \, g^\prime_{NN} = 0.7, \, g^\prime_{N \Delta} = g^\prime_{\Delta N} = 0.5, \, g^\prime_{\Delta\Delta} = 0.5 $$ 
The inclusive RPA responses functions are deduced from 
the corresponding inclusive RPA polarization propagators by the usual relation:
\begin{equation} \label{eq:6}
R (\omega,q) = -\frac{1}{\pi} \, \mathrm{Im}( \Pi(\omega,q,q) )
\end{equation}
We perform the calculation of the partial RPA responses as follows. Starting 
from the RPA equation (\ref{eq:4}), we write the imaginary part of 
$ \Pi(\omega,\gvec{q},\gvec{q^\prime}) $ in the following form:
\begin{equation} \label{eq:7}
\mathrm{Im}( \Pi ) = | 1 + \Pi \, V |^2 \, \mathrm{Im}( \Pi^0 ) \, + \, 
| \Pi |^2 \, \mathrm{Im} V 
\end{equation}
with $ \mathrm{Im}( \Pi^0 ) = \sum_{k=1}^{n_k} \mathrm{Im}( \Pi^0(k) ) $.
This sum rule gives the different contributions to the inclusive RPA response 
functions. The first terms in eq. (\ref{eq:7}) are reminiscent of the bare 
case. Indeed we recognize the bare partial response functions (apart from the 
trivial $ -\pi $ factor) corrected by a factor involving the inclusive RPA 
polarization propagator and the effective interaction. The partial RPA 
response functions, defined by:
\begin{equation} \label{eq:8}
R(k)(\omega,q) = - \frac{1}{\pi} \, | 1 + \Pi \, V |^2 \, \mathrm{Im}( 
\Pi^0(k)(\omega,q,q)) 
\end{equation}
are represented by the graphs (a) to (d) on fig. (\ref{fig:2}), where the 
hatched rings correspond to the inclusive polarization propagator solution of eq. (\ref{eq:4}), the non hatched rings to the bare partial polarization propagators (the dotted line means that 
we take the imaginary part of these propagators) and the dashed lines to the 
effective interaction. It is easy to recover on these graphs the different 
terms of the development of eq. (\ref{eq:8}). Note that in the RPA case, a 
$ PP^\prime $ reaction channel ($ P,P^\prime \, = \, N,\Delta $) gets 
contributions from every $ QQ^\prime $ configurations ($ NN, \, N\Delta, \, 
\Delta N, \, \Delta\Delta $). The last term in eq. (\ref{eq:7}) corresponds to 
the "coherent" response function:
\begin{equation} \label{eq:9}
R_{coh}(\omega,q) = -\frac{1}{\pi} \, | \Pi |^2 \, \mathrm{Im}V
\end{equation}
It is absent of the response spectrum when the effective interaction is 
switched off. In the domain of energy considered here the sole contribution to 
this channel comes from the pion exchange. This process corresponds to the 
emission of a pion on its mass-shell, the nucleus remaining in its ground 
state. It is represented by the graph (e) on fig. (\ref{fig:2}) where the dashed line stands for the exchange of a pion. The 
implications of these partial reaction channels will be discussed in the 
following sections.

\section{Cross sections}     

The next step is the calculation of the neutrino-oxygen cross section. The 
doubly differential cross section is given in a first approximation by eq. 
(\ref{eq:1}). Our final calculation relies on a more complete expression, 
which we will briefly describe in the following, but the main features remain 
unchanged. First it is essential to note that the neutrino-nucleus reaction is 
strongly dominated by the transverse spin-isospin channel. This is clear from 
eq. (\ref{eq:1}). Indeed the terms multiplying the transverse responses, 
depending on the axial and magnetic form factors, have a much larger magnitude than for the longitudinal case. 
Furthermore the $ NN $ quasi-elastic spin-isospin longitudinal response is totally suppressed in the cross section. This suppression arises 
from an exact cancellation at the top of the quasi-elastic peak between the 
various terms entering the contraction of the leptonic and the hadronic 
tensors. This suppression is only partial in the Delta resonance or in the "dip" region (the region intermediate between the quasi-elastic and the Delta peaks). 
This result is in contradiction with the study of ref. \cite{oset/singh} where the relative 
weights of the transverse to the longitudinal responses were assumed to be 2:1. 
Note however that the suppression of the $ NN $ longitudinal response is no 
more exact when one considers the complete expression of the doubly differential 
cross section, which includes the terms involving the charged lepton 
mass (in fact we consider only the muon mass) and the terms up to order 
$ (p/M)^2 $ in the reduction of the hadronic current. The contributions of the 
$ NN $ longitudinal response are then of order $ (m_l/M)^2 $, where $ m_l $ 
denotes the mass of the charged lepton, and of order $ (p/M) $. These 
corrections are rather weak. As another source of corrections we have also considered the renormalization of the axial charge by the mesonic exchange currents, because the suppression of the $ NN $ longitudinal response involves the time component of the axial current. Following the parameterization of ref. \cite{warburton2} we make the replacement \mbox{$ g_A \longrightarrow g_A ( 1 + \delta ) $} in the time component of the axial current. The contribution of the $ NN $ 
longitudinal response is then of order $ \delta^2 $. Even with the relatively 
high value $ \delta \sim 0.5 $ suggested in ref. \cite{towner}, the contribution of the 
$ NN $ longitudinal response remains weak. The same conclusion holds for the 
$ N\Delta $ and $ \Delta\Delta $ longitudinal responses which are widely 
dominated by the corresponding transverse ones. Note that in the 
antineutrino-nucleus reactions the weight of the transverse channel is 
somewhat reduced because of the change in sign in the interference term. 
However even in this case, the transverse responses correspond to ~75 \% of 
the total, the remaining arising essentially from the $ NN $ pure isospin response. 

\begin{figure}[ht]
\begin{center}
\includegraphics[width=12cm,height=10cm]{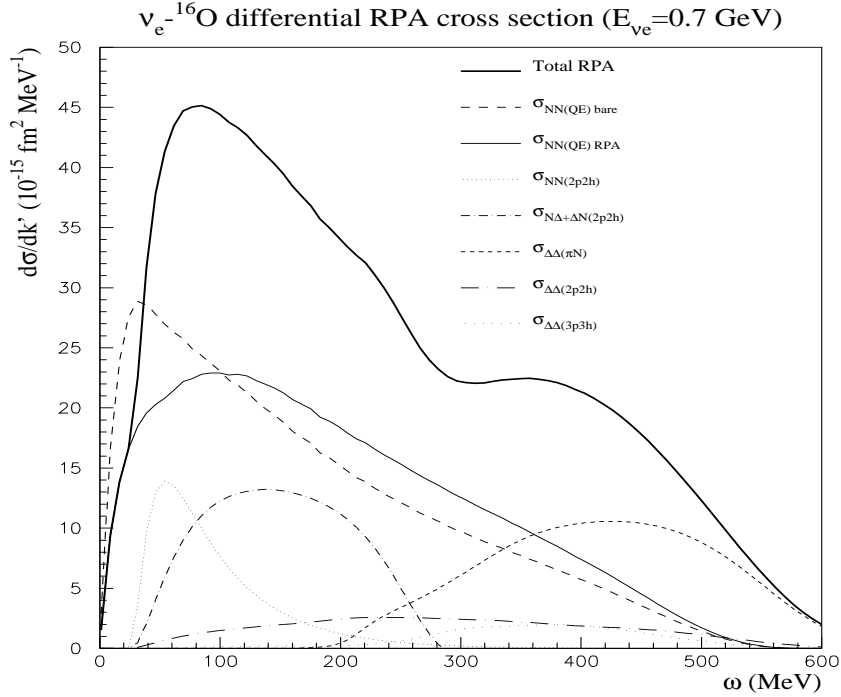}
\caption{\label{fig:3} \textit{Differential charged current 
$ \nu_e- ^{16}\hbox{O} $ interactions cross-section versus the energy transfer. 
The thick curve represents the inclusive cross-section. The following exclusive 
contributions to the inclusive cross-section are displayed: 
$NN$ (quasi-elastic) (full thin curve), $NN$ (2p-2h) (short dotted curve), 
$N\Delta + \Delta N$ (2p-2h) (short dot-dashed curve), $\Delta\Delta$ 
($ \pi N $) (short dashed curve), $\Delta\Delta$ (2p-2h) (long dot-dashed curve) 
and $\Delta\Delta$ (3p-3h) (long dotted curve). Also shown is the "bare" 
$NN$ quasi-elastic cross-section (long dashed curve).}} 
\end{center}
\end{figure}

We will now investigate the implications of these global features on the 
simply differential cross section \mbox{$ \partial\sigma / \partial k^\prime $},
which is obtained from the doubly differential cross section by a numerical 
integration over the solid angle. The great interest of our method is the 
separation of the inclusive cross section on partial contributions. This 
separation is simply achieved by the replacement, in the expression of the 
cross section, of the inclusive response functions with the "exclusive" ones, 
calculated in the previous section. The results are shown on fig. (\ref{fig:3}) 
which displays the simply differential cross section versus the energy transfer, fixing for the sake of illustration a neutrino energy of 700 MeV.        
  
The inclusive cross section is given by the thick curve. It gets its main 
contribution from the $ NN $ quasi-elastic channel (thin full line) which 
peaks at relatively low energy transfer. For the sake of comparison we have 
shown the contribution of the $ NN $ quasi-elastic channel without RPA (thin 
long dashed line). We observe that the cross section is reduced and hardened 
in the RPA case. This result is in full agreement with that of ref. \cite{engel/vogel}. The 
shift in strength reflects the dominance of the transverse response. Indeed 
the Landau-Migdal interaction is repulsive for all values of the transfer and 
the $ \rho-\mathrm{exchange} $ piece is not attractive enough in the domain of 
energy considered here to counteract this feature. This repulsive effective 
interaction hardens and reduces the transverse response functions. For instance this conclusion no longer holds in the longitudinal channel where 
the $ \pi-\mathrm{exchange} $  is attractive enough to overcome the 
Landau-Migdal interaction and to create a collective mode (the so-called 
\textit{pionic branch} [5,6]). But, as we mentioned previously, the 
suppression of the longitudinal channel makes the neutrino a poor probe of 
this pionic mode. 

The effect of the RPA correlations are less strong in the others channels and 
are not shown here. The $ \Delta\Delta $ ($ \pi N $) channel (short dashed curve on the figure) arises at high energy transfer ($ \omega \sim 450 $ MeV). This is in good agreement with the 
result of the relativistic calculations given in ref. \cite{kim/schramm/horowitz} where the Delta was taken into account as a free resonance and where the RPA correlations did not 
include Delta-hole configurations. The agreement between the two calculations 
is not surprising. Indeed we use relativistic kinematics for the evaluation of 
the polarization propagators as in ref. [5,6] and we include the 
terms up to second order in the $ (p/M) $ reduction of the hadronic current as 
mentioned above. Furthermore the RPA effects on the transverse response in 
the Delta region, that is at high transfer, are somewhat reduced in the cross 
section by the form factors and the differences between RPA and bare 
transverse response functions are very weak at these values of the transfer. 
Thus our calculations show that a free Delta resonance gives a rather good 
approximation of the $ \Delta\Delta $ ($ \pi N $) (or "quasi-elastic") channel. 
This result corroborates the fact that the "quasi-elastic" Delta width in the 
nuclear medium is close to the its free value, the Pauli blocking being 
cancelled by the other mechanisms taken into account. 
 
The most interesting feature of the cross section is the importance of the 
(\textit{np-nh}) channels. The kinematics of the neutrino-nucleus reaction 
tends to favor the  $ NN $ (\textit{2p-2h}) channel (short dotted curve) which peaks at low energy transfer. However the $ N\Delta + \Delta N $ (\textit{2p-2h}) channel (short dot-dashed curve) gives a rather large contribution to the inclusive cross section and has an extended 
spectrum in the "dip" region. Its importance has been pointed out in the  $ (e,e^\prime) $ scattering 
where it is necessary to reproduce the experimental data in the "dip" region 
(for example see ref. \cite{alberico/ericson/molinari}). Finally note that the $ \Delta\Delta $ 
(\textit{2p-2h}) (long dot-dashed curve) and (\textit{3p-3h}) (long dotted curve) spectra extend over a wide range of energy transfer, while the $ \Delta\Delta $ ($ \pi N $) channel is concentrated in the so-called Delta peak. They give a little contribution to the inclusive 
cross section (in particular the (\textit{3p-3h}) channel is rather weak) but 
the extension of their spectra will have important consequences in the specific events yields.

The results obtained in this section show that the inclusive neutrino-oxygen cross 
section is strongly modified with respect to the free  $ NN $ quasi-elastic 
case, which is quite often the sole channel entering into the calculations. In 
particular we have seen the occurrence of large contributions from the two- and 
three-body channels. The main effect of the nuclear correlations is the 
hardening of the $ NN $ quasi-elastic channel. They have rather low impact on 
the others reaction channels and therefore could be legitimately neglected. 

\section{Events yields}     
     
\begin{wrapfigure}[24]{r}{9 cm}
\begin{center}
\includegraphics[width=8cm,height=7cm]{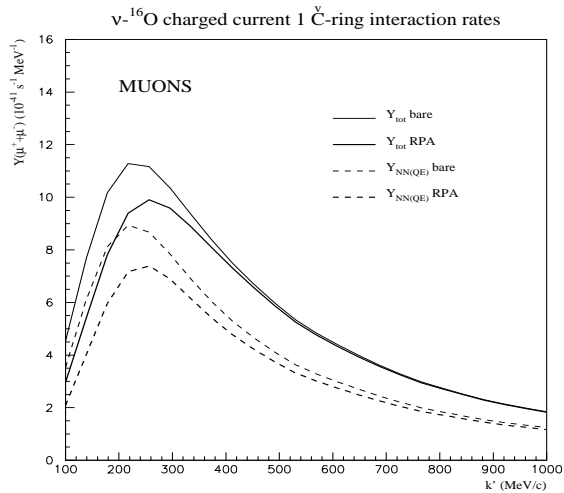}
\caption{\label{fig:4} \textit{One \v{C}erenkov $ (\nu_\mu + \bar{\nu}_\mu)- ^{16}\hbox{O} $ events yields versus the muon momentum. The full curves correspond to the total 1 \v{C}.R. events yields with (thick curve) and without (thin curve) RPA. The dashed curves correspond to the $ NN $ quasi-elastic 1 \v{C}.R. events yields with (thick curve) and without (thin curve) RPA.}} 
\end{center}
\end{wrapfigure}

In this section we compute numerically the neutrino-oxygen events yields for a 
fixed charged lepton momentum:
\begin{equation} \label{eq:10}
Y(\nu_\alpha + \bar{\nu}_\alpha)(k^\prime) = 
\int_{E_{k^\prime}}^{\infty} \, dE \, \left( 
\Phi_{\nu_\alpha}(E) \, \frac{\partial\sigma}{\partial k^\prime}(E,k^\prime) 
\Phi_{\bar{\nu}_\alpha}(E) \, \frac{\partial\bar{\sigma}}{\partial k^\prime}
(E,k^\prime) \right) 
\end{equation}
where $ E $ is the neutrino energy, $ \Phi_\nu $ ($ \Phi_{\bar{\nu}} $) the 
incoming neutrinos (antineutrinos) flux 
and $ \partial\sigma/\partial k^\prime $ ($ \partial\bar{\sigma}/\partial 
k^\prime $) the neutrino-oxygen (antineutrino-oxygen) cross section 
computed in the previous section. We use the fluxes of Bartol \cite{barr/gaisser/stanev} in our 
calculations for the sake of comparison with the results of ref. \cite{engel/vogel}. The main feature of these fluxes is their sharp decrease with increasing neutrino energy. Note that several theoretical attempts have been undertaken to compute these atmospheric neutrinos fluxes. The sources of possible differences between three models have been analyzed in ref. \cite{gaisser/honda/etal}. The predictions of these models on the flavor ratio agree at a $ \sim $ 5 \% degree of accuracy. Anyway the divergences in the absolute fluxes remain 
rather large ($ \sim $ 20 \%). Furthermore new measurements on the primary 
cosmic rays fluxes could lead to some modifications with respect to the 
present situation. The cumulated uncertainties on the neutrino fluxes and on 
the neutrino-oxygen cross sections could lead to modifications in the 
experimental analysis, even if they remain unlikely to explain the atmospheric 
neutrinos anomaly.
To perform an analysis of the events yields, we need to classify the partial 
reaction channels according to the number of \v{C}erenkov ring(s) they produce. 
This classification has been elaborated within a few rough assumptions. First 
we consider that every nucleon ejected from the nucleus remains under 
\v{C}erenkov threshold. In water, the threshold kinetic energy for a particle 
of mass  $ m $ is \mbox{$ E \sim 0.5 \, m $}. For a nucleon, produced through Delta 
decay or (\textit{2p-2h}) mechanisms, the assumption is fairly good. On the 
opposite, we assume that every pion which escapes the nuclear medium produces 
a \v{C}erenkov ring. The threshold energy for a pion being $ \sim $ 70 MeV, 
this assumption is believed to be reliable. Then the partial reaction channels 
leading to one \v{C}erenkov ring, in charged current interactions, are the  
$ NN $ quasi-elastic one, which is usually taken into account in the 
simulations, and the (\textit{np-nh}, n$ > $1) type channels (both $ NN $, $ N\Delta $, 
$ \Delta N $ and $ \Delta\Delta $). The remaining reaction channels, 
$ \Delta\Delta $ ($ \pi N $) and coherent pion production, are supposed to 
lead to at least two \v{C}erenkov rings. The results for the 1 \v{C}.R. events 
yields, which are relevant in the atmospheric neutrino experiments, are shown 
on fig. (\ref{fig:4}) for incident $ \nu_\mu $ and $ \bar{\nu}_\mu $, the full 
curves corresponding to the total 1 \v{C}.R. events yields and the dashed 
curves to the sole  $ NN $ quasi-elastic 1 \v{C}.R. events yields. We give the 
results of the calculations without (thin curves) and with (thick curves) RPA. 

First we observe that the RPA tends to reduce the events yields. This is not 
hard to understand. Indeed the RPA tends to harden the cross sections, 
\textit{i.e.} to push the strength towards higher energies. But the fluxes 
decrease with increasing energies and therefore the higher energies are 
disfavored. This reduction affects mostly the $ NN $ quasi-elastic channel in 
accordance with the result obtained for reaction cross sections. The maximal 
reduction factor is of the order of 10 \%. A more interesting feature is the 
strong enhancement of the absolute events yields implied by the 
(\textit{np-nh}) channels. At the maximum value of the yields, the enhancement 
of the total yield with respect to the $ NN $ quasi-elastic one is around 
30 \%. This result reflects the main features of the cross sections. 
Furthermore one must be aware that this result is a lower limit of the true enhancement. 
Indeed we know that pions can be re-absorbed in the nucleus. 
Therefore the events produced in the $ \Delta\Delta $ ($ \pi N $) channel can 
also lead to one \v{C}erenkov ring if the pion does not escape from the 
nucleus. 
Thus we can conclude that the RPA 1 \v{C}.R. events yields induced by 
charged current interactions is globally enhanced with respect to the $ NN $ 
quasi-elastic 1 \v{C}.R. events yields without RPA. 
The difference between the 
two calculations could be responsible for the small discrepancy between the 
experimental and simulated events distributions in Super-Kamiokande 
\footnote{We thank Y. Declais for attracting our attention on this point.}. 
But we need complementary informations to ensure this conclusion. Indeed it is hard to establish the enhancement factor firmly. Our present analysis leads 
to a maximum enhancement factor of the order of \mbox{$ \sim $ 20 \%}. We have 
already mentioned that the problem of pion absorption, which is not yet considered in our calculations, could still enhance this factor. We should also be aware that there exists some misidentification problems which could have a more or less large effect. One of the 
misidentification source, pointed out by the authors of ref. \cite{kim/schramm/horowitz}, is the 
"coherent" pion production. Their analysis is based on the assumption that the 
forward peaked angular distribution of the coherent pions entails the coherent 
pions to be emitted with a small angle with respect to the charged lepton 
direction. This could mimic a "shower" which could be interpreted as an $ e-\mathrm{type} $ 
event, whatever may be the flavor of the incoming neutrino. Our calculations 
show that this coherent channel brings a tiny contribution (less than 2 \% of 
the total $ \nu_e + \bar{\nu}_e $ events yield) which makes it irrelevant in 
the atmospheric neutrino anomaly. The suppression of this channel is understandable. Part of it is due to the nuclear form factors effects as discussed for example in \cite{ericson/weise}. In addition the coherent response manifests itself mainly in the longitudinal spin-isospin channel and we have seen that this channel is suppressed with respect to the transverse one in the 
neutrino-nucleus reactions. The coherent pions should not be a problem. 

The case of the neutral currents is less clear. In charged current interactions, 
pions ("coherent" or not) lead to, at least, 2 \v{C}.R. events and are excluded 
from the analysis. But in neutral currents interactions they lead to 1 \v{C}.R. 
events, because the scattered neutrino does not produce any ring. We have computed the neutral current events yields in each reaction channel, the few differences with respect to the charged current case being easily included in the formalism. The problem arising then is the classification of these $ \pi-\mathrm{like} $ 1 \v{C}.R. events. Indeed in absence of indication on the experimental $ \pi/\mathrm{lepton} $ discrimination efficiency in the water \v{C}erenkov experiments, it is not possible to draw firm conclusions on the role played by the neutral currents. This problem has to be investigated further (for more details on the results of the calculations, see \cite{thesis}).  

Finally we study the evolution of the flavor ratio with the charged lepton 
momentum:
\begin{equation} \label{eq:11}
R_{\mu/e}(k^\prime) = Y( \nu_\mu + \bar{\nu}_\mu )/Y( \nu_e + \bar{\nu}_e ),
\end{equation}
where $ Y $ denotes the events yields defined by eq. (\ref{eq:10}). We compare the total 1 \v{C}.R. events yields flavor ratio with the $ NN $ quasi-elastic 1 \v{C}.R. events yields flavor ratio. The result is shown in table (\ref{table:1}) where the ratio of ratios has been calculated for four relevant lepton momentum. There is almost no modification between the two situations. This conclusion strengthens the usual assumption that uncertainties due to nuclear effects cancel when one considers ratios of events rates. Here the maximum effect on the flavor ratio is less than \\ \mbox{10 \%.}

\begin{table}[ht]
\begin{center}
\begin{tabular}{|c||c|}
\hline
$ k^\prime $ (MeV/c) & $ R_{\mu/e}(NN \, q.e.) / R_{\mu/e}(Total) $ \\ 
\hline 
100 & 1.060 \\ 150 & 1.040 \\ 250 & 0.999 \\ 400 & 1.001 \\
\hline
\end{tabular}
\end{center}
\caption{\label{table:1} \textit{Comparison of the total and $ NN $ 
quasi-elastic 1 \v{C}.R. events yields ratios for four lepton momentum.}}  
\end{table}

\begin{wrapfigure}[23]{r}{9 cm}
\begin{center}
\includegraphics[width=8cm,height=8cm]{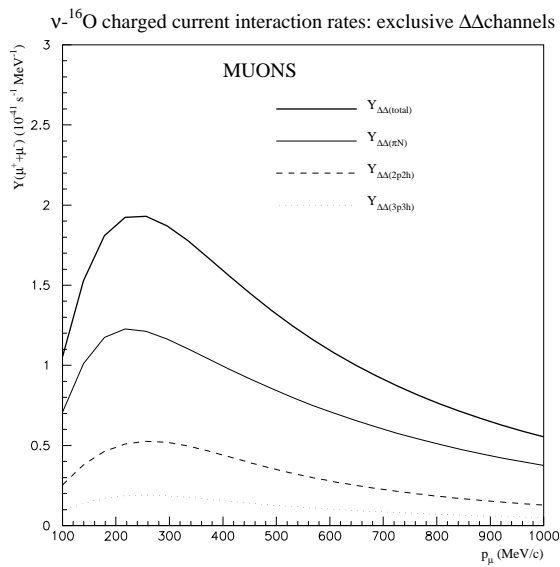}
\caption{\label{fig:5} \textit{Contributions to the total $ \Delta\Delta $ 
events yield (full thick curve) of the partial channels: ($ \pi N $) 
(full thin curve), (2p-2h) (dashed line), (3p-3h) (dotted line).}} 
\end{center}
\end{wrapfigure}

We conclude this work by mentioning the problem of pion emission in neutrino-oxygen 
interactions. On one side we have shown that the cross sections of the $ \Delta\Delta $ (\textit{2p-2h}) and (\textit{3p-3h}) partial channels, which do not not lead to pion emission (\textit{non pionic} channels), extend over a broad region in transfer energy, while the \textit{pionic} channel $ \Delta\Delta $ ($ \pi N $) is peaked at high transfer energy (see fig. (\ref{fig:3})). On the other side the neutrino flux lowers the weight of the high energies and favors the low energy components of the spectrum. Then the pionic $ \Delta\Delta $ channel will be more suppressed by the incident neutrino flux than the non pionic one. This result is shown on fig. (\ref{fig:5}) where the total $ \Delta\Delta $ events yield (full thick curve) is split into its three contributions: ($ \pi N $) (full thin curve), \textit{2p-2h} (dashed curve), \textit{3p-3h} (dotted curve) in the case of $\mu-\mathrm{type}$ events. 

The main result is that the fraction of the non pionic channels over the pionic 
one is around 50 \%. This result remains valid for every values of the lepton momentum. Thus the  (\textit{np-nh}) excitations play an important role in the events yields although their reaction cross sections are relatively low. Finally we would like to point out that some pion production mechanisms, which do not reduce to a simple response function, are still absent of our formalism. For example we have omitted in the production through the vector current, the Kroll-Ruderman and the pion-in-flight terms which play an important role. Improvements to our present calculations are in progress. Nevertheless these limitations of our present calculations do not alter the need of including the effects of the partial $ \Delta\Delta $ (\textit{np-nh}) reaction channels to avoid an overestimation of the number of pions effectively produced in the neutrino-oxygen interactions.

\section{Conclusion}     
     
In this work we have studied the effects of nuclear correlations on the charged current neutrino-oxygen cross sections and events yields in specific exclusive reaction channels. We have shown that besides the quasi-elastic channel the (\textit{np-nh}, n=2,3) excitations also lead to one \v{C}erenkov ring events, which are retained for the analysis of the experiments using large underground water \v{C}erenkov detectors. The enhancement in the one \v{C}erenkov ring events yields is large and could still be increased when some processes, such as pion absorption in nuclei or neutral currents events, are taken into account. It is therefore important to take these nuclear effects into account to perform a calculation of absolute events yields. We have also shown that the flavor ratio $ R_{\mu/e} $ is not significantly altered. 
We have applied our model to others neutrino-nucleus reactions. In particular we have studied the case of iron which is the target-nucleus in the neutrino experiments using calorimeters. The conclusion on the cross sections are the same than the one presented here in the case of oxygen. However such experiments measure more detailed observables than the water \v{C}erenkov detectors, like the energy and momentum spectra of the particles in the final state. The description of these experiments requires the extension of our model. The present work already shows the necessity of taking into account nuclear correlations involving multi-nucleon excitations.

\section*{Acknowledgements}
          
We wish to gratefully acknowledge J. Delorme for his numerous contributions to this work. We are indebted for enlightening discussions and critical reading of the manuscript to J. Delorme, M. Ericson and G. Chanfray. We also thank Y. Declais, P. Lipari and S. Katsanevas for stimulating discussions.

\end{document}